\documentclass[aip,rsi,reprint]{revtex4-1}

\usepackage{hyperref}
\usepackage[utf8]{inputenc}
\usepackage[english]{babel}
\usepackage{amsmath}
\usepackage{graphicx}
\usepackage[labelformat=simple]{subcaption}

\begin{document}

\title{Non-blocking programmable delay line with minimal dead time and tens of picoseconds jitter}

\author{Glib Mazin}
\affiliation{Department of Optics, Faculty of Science, Palack\'y University, 17.\ listopadu 12, 77146 Olomouc, Czechia}

\author{Ale\v{s} Stejskal}
\affiliation{Department of Experimental Physics, Faculty of Science, Palack\'y University, 17.\ listopadu 12, 77146 Olomouc, Czechia}

\author{Michal Dudka}
\affiliation{Department of Optics, Faculty of Science, Palack\'y University, 17.\ listopadu 12, 77146 Olomouc, Czechia}

\author{Miroslav Je\v{z}ek}
\email{jezek@optics.upol.cz}
\affiliation{Department of Optics, Faculty of Science, Palack\'y University, 17.\ listopadu 12, 77146 Olomouc, Czechia}


\begin{abstract}
We report a non-blocking high-resolution digital delay line based on an asynchronous circuit design. Field programmable gate array logic primitives were used as a source of delay and optimally arranged using combinatorial optimization. This approach allows for an efficient trade-off of the resolution and a delay range together with minimized dead time operation. We demonstrate the method by implementing the delay line adjustable from 23~ns up to 1635~ns with a resolution of 10~ps. We present a detailed experimental characterization of the device focusing on thermal instability, timing jitter, and pulse spreading, which represent three main issues of the asynchronous design. We found a linear dependence of the delay on the temperature with the slope of 0.2~ps$\cdot$K${}^{-1}$ per a logic primitive. We measured the timing jitter of the delay to be in the range of 7~ps -- 165~ps, linearly increasing over the dynamic range of the delay. We reduced the effect of pulse spreading by introducing pulse shrinking circuits, and reached the overall dead time of 4~ns -- 22.5~ns within the dynamic range of the delay. The presented non-blocking delay line finds usage in applications where the dead time minimization is crucial, and tens of picoseconds excess jitter is acceptable, such as in many advanced photonic networks.
\end{abstract}
\maketitle

\section{Introduction}

Digital delay line (DDL) is essential tool in experimental physics research and engineering applications. DDLs represent core building blocks in nuclear physics instrumentation, laser synchronization, and photonic networks, where a target delay ranges from nanoseconds to microseconds with sub-nanosecond precision. Tunable delays up to a few microseconds are often needed in correlation measurements of cascade decays \cite{Marques2014}. Electronic delays with tens of picoseconds resolution and dynamic range of hundreds nanoseconds are necessary to compensate for the length fluctuations of optical communication channels \cite{Sliwczynski2011,Boaron2018}. DDLs are frequently utilized in photonic quantum technology \cite{Sciarrino2019,Pryde2019} for photon coincidence detection \cite{Zhang2016, Hlousek2019,Arabul2020} and switching \cite{OBrien2016,Xiong2016,Kaneda2019,Svarc2020}. These applications require synchronization of several single-photon detectors and active components, like modulators or switches. Delays from tens to hundreds of nanoseconds are needed with precision of tens or hundreds of picoseconds, which corresponds to timing jitter of photonic detectors. Single-photon avalanche diodes may serve as an example of frequently used photonic detectors with 50-300 ps jitter and 10-50 ns dead time (typical values). DDLs should have comparable or lower jitter and dead time to be efficiently combined with these detectors.

DDLs are generally divided into two main groups: blocking and non-blocking. The blocking design typically utilizes system clock as a time reference, counts of which determine the amount of introduced delay between input and output signal. Resolution of this method is constrained to the clock period, however, various sub-clock techniques allow to go beyond the basic clock resolution, e.g. multiple clock signals with introduced relative phase shift or Vernier delay lines \cite{Dudek2000,Zhu2017}. The main disadvantage of this approach is a dead time of the delay line which can be understood as a time of inability to react or process any input signal due to processing of the previous input. This effect occurs naturally due to the usage of the reference clock, and has a value equal to the introduced delay $n \cdot T_{clk}$, where $n$ is the number of time intervals and $T_{clk}$ is the period of reference clock respectively \cite{Zhu2017}.
The dead time reduces DDL throughput and affects statistics of the signal \cite{Bedard1967,Straka2020}, which significantly limits applications of blocking DDLs in quantum technology.

The non-blocking design typically exploits inevitable intrinsic fixed delay of electronic components or primitives used in construction. One of the most used circuit designs here is so-called tapped delay line, where components are connected serially. Consequently, the signal is extracted from output of the corresponding tap and thus with certain amount of propagation delay. Tapped delay lines typically produce short delays and are often used in combination with synchronous delays.

\begin{figure}[ht] \centering
	\includegraphics[width=1.0\columnwidth]{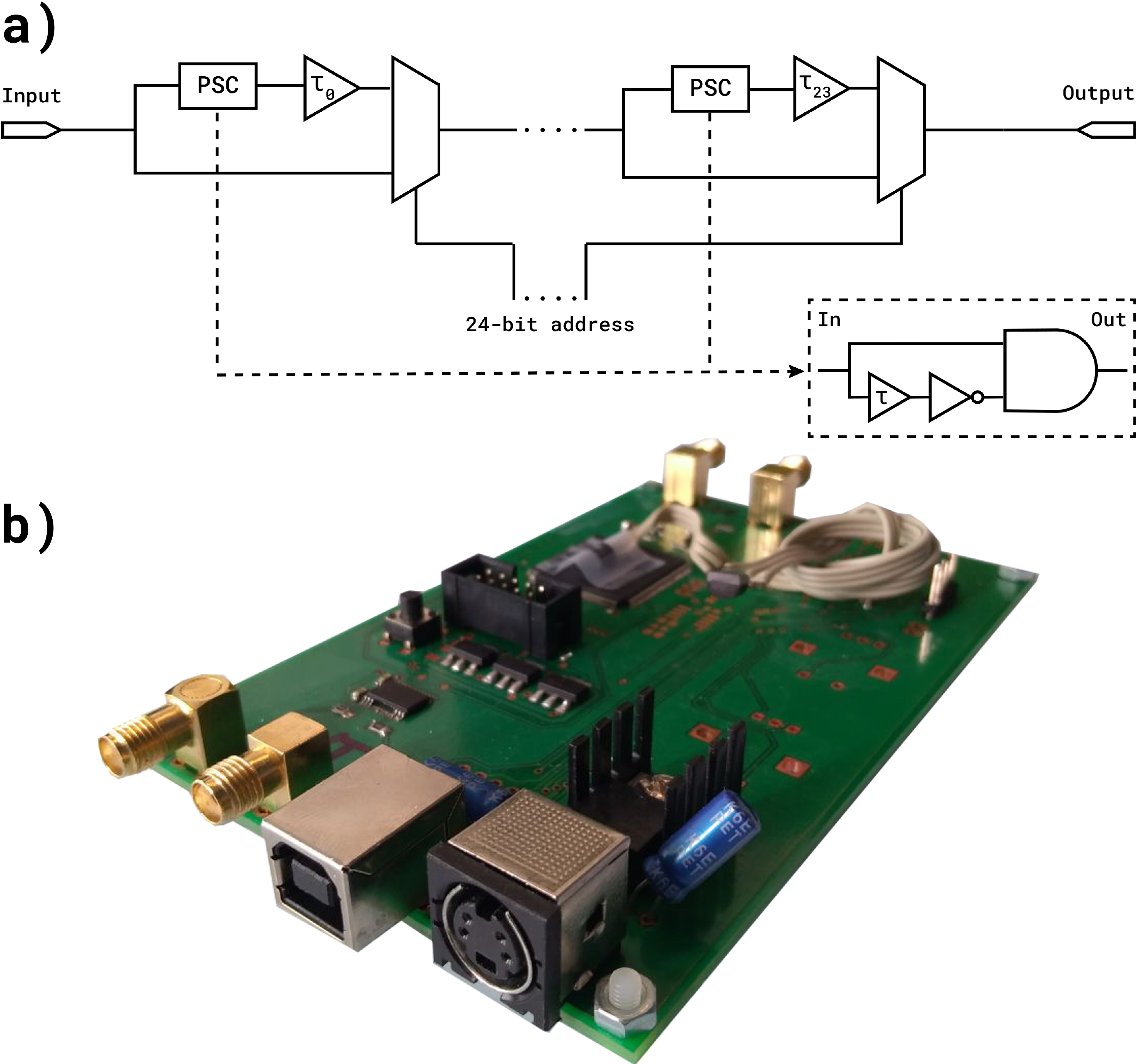}
	\caption{a) Scheme of the presented delay line: individual constituent delay blocks $\tau_n$ are switched by
multiplexers (MUXs) to match optimally the target delay. Pulse shrinking circuits (PSCs)
are employed to reduce pulse dispersion, see text for details.
 b) Photo of the developed device.}
	\label{fig:scheme}
\end{figure}

Traditional platforms used for DDLs are discrete components and microcontroller units. Although, field programmable gate arrays (FPGAs) are becoming more and more utilized in the area of digital delay generators. The main disadvantage of using discrete components is their limited tunability without intervention at hardware level. Microcontrollers offer programability and a wide range of the generated delays up to hundreds of seconds, and granularity in order of tens of nanoseconds \cite{Jang2007,Handa2007,Eyler2011,Hosak2019}. However, they are clock-based, i.e. dead-time limited, and not convenient for sub-nanosecond delay generation. FPGAs offer unique reprogramming flexibility, multi-channel usage and easy implementation of synchronous as well as asynchronous structures in the designs. Many custom-built DDLs were demonstrated exploiting combinations of discrete components and FPGAs \cite{Yue2011} or solely FPGAs \cite{Szplet2015,Savaria2017,Zhu2017}. Recent study shows remarkable resolution of units of picoseconds, however, still it is a dead-time limited solution \cite{Zhang2019}. Consequently, we lack a non-blocking solution with both wide dynamic range and fine granularity.

Here, we propose a novel design of a non-blocking delay line based on the switching between optimally selected variations of a few precalibrated delay blocks in a FPGA. We carefully characterize the performance of the presented device. Particularly, we present the measurement of its temperature dependence, timing jitter, and pulse dispersion as these parameters are crucial for the adopted asynchronous design. The developed delay line provides a programmable delay up to 1.6 $\mu$s with the resolution of 10 ps and the maximum dead time of 22.5 ns. In envisioned applications, where high-rate randomly arriving signals from optical or particle detectors are delayed, it provides minimal excess dead time and features timing jitter lower or comparable to commonly used detectors.


\section{Design}
\label{Sec:design}

Our aim is to utilize logic primitives of the FPGA to construct a delay line in such a way that it will guarantee both wide dynamic range and high resolution. The delay line is designed as a sequence of serially connected 24 stages. Each stage contains a delay introducing element and a switching element, which enables to switch between delay stages and thus to chose the desired delay value. In our case, the delay introducing element is an array of LUTs and the switching element is 2-to-1 multiplexer (see Fig. \ref{fig:scheme} a).
The route in each stage splits into two, one delaying the input signal, the other propagating the signal with just an intrinsic delay of the multiplexer (zero-delay path). The intrinsic delays contribute to the minimum delay that can be reached by the delay line when all stages are set to zero-delay path.
To provide fine granularity for shorter delays, each pair of the first 10 stages is designed to have the same number of LUTs (i.e. 1, 1, 2, 2, $\ldots$, 5, 5), since the FPGA manufacturing deals with physical parameters variation and each LUT may have slightly different propagation delay. The number of LUTs in the remaining 14 stages forms a geometric series with geometric factor of approximately 1.6 except for the last stage, in which the number of LUTs was limited to 1729 by the maximum available LUTs of the employed FPGA.

Delays introduced by the individual stages are measured during the calibration of the DDL by the same method used for characterization of the whole device as described in Section \ref{Sec:temporal_characterization}.
The measured values compose a $1\times24$ vector $(\tau_0, \ldots, \tau_{23})$. We calculate delays of all possible $2^{24}$ stage variations as a matrix product of the measured values and $24 \times 2^{24}$ stage combinations binary matrix:
\[
\begin{pmatrix} 
\tau_0, & \!\ldots, & \!\tau_{23}\\
\end{pmatrix}
\begin{pmatrix} 
0 & 1 & 0 \dots 1\\
0 & 0 & 1 \dots 1\\
\vdots & \vdots & \vdots \dots \vdots\\
0 & 0 & 0 \dots 1\\
\end{pmatrix}
=
\begin{pmatrix} 
\tau_0, & \!\ldots, & \!\tau_0 + \cdots + \tau_{23}\\
\end{pmatrix}
\]
Sorting the obtained vector of variations in ascending order, taking differences of each pair of its values i.e. $\Delta t_n = t_{n+1} - t_n$ and performing a selection, the vector of delay step values with target granularity can be distilled. The selection algorithm represents a simple linear search through the array of differences $\Delta t$ with a condition which matches the target granularity value. As a result, $160\times10^3$ matched delay values with granularity $10 \pm 5$ ps and corresponding binary control words are loaded in the DDL to generate a required delay value. 
The key point of the proposed design is a tunable trade-off between high dynamic range and fine granularity. Although we picked such values that satisfy requirements of our experiment, i.e. the delay range from 23 ns to 1635 ns and granularity 10~ps, one can modify the design according to one's needs.

Propagating through a sequence of LUTs results in non-negligible dispersion, which will be discussed in Section \ref{Sec:temporal_characterization}. The pulse dispersion causes spreading of the propagating pulses, and may negatively affect effective dead time of the DDL, i.e. two pulses with a small temporal separation might not be resolved at the output of the device.
To diminish this issue, we included six pulse shrinking circuits (PSCs): at the beginning of the DDL and before the last five stages. The first PSC performs time width standardization and makes input pulses to be of the same length. The similar standardization can be performed at the output of the DDL.
The five PSCs before the last five stages (the largest delays) combat the pulse dispersion by resetting the pulse length to 1 ns, thus ensure constant output pulse length and decrease DDL dead time. Adding PSCs does not affect the calibration of the device as the delays of individual stages are measured including the PSCs; the same holds for delays of the multiplexers.


\section{Implementation}

The developed DDL was implemented in Cyclone IV EP4CE6E22C6N FPGA from Altera on a custom-built electronic board. Digital circuit was designed by means of VHDL using the Intel Quartus Prime Lite edition software. Communication of the device with PC via USB to serial converter is performed by STM32F334K8 microcontroller. It also stores selected control words in
AT25SF161 16Mbit FLASH memory and controls other parts of the board including the DDL. Input analog signal is converted to digital by LTC6754 fast LVDS comparator to match the FPGA input. The delayed output from FPGA is shifted by SN74LVC4245A voltage level shifter to +5 V level to be compatible with other laboratory devices. Current operation mode is single-channel with straightforward multi-channel extension possibility.

Input signal is connected to the board via SMA connector and is terminated with 50 Ohm. Maximum allowed input voltage ranges from -0.2 V to +5.1 V. However, range from 0 V to +3.3V is recommended because comparator threshold can be set in this range by the 12-bit digital-to-analog converter which is incorporated in the microcontroller. The comparator contributes to zero-delay path with maximally 1.6 ns considering sufficient overdrive higher than 115 mV\cite{LTC6754}. The LVDS signal from comparator is received by FPGA and goes through digital delay line described in Section \ref{Sec:design}.

Multiplexers of the DDL are directly controlled by the FPGA-implemented 24-bit shift register, which is loaded externally by the microcontroller with the specified control word. When the signal passes the zero-delay path, it goes through 25 LUTs (input PSC 1 LUT, multiplexers in all stages 24 LUTs), LVDS input buffer, and output buffer. This propagation delay through FPGA is the most significant part of the zero-delay path (ca. 14-16 ns). The average delay per LUT in individual stages, which were measured during calibration, appeared to be dependent on number of LUTs used in the stage. For lower number of LUTs less than 10 the average delay per LUT varies within the range of $350 \pm 250$ ps per LUT. For larger delay stages (100 LUTs and more) this value converged to 270 ps per LUT. This can be the consequence of that the signal routing and LUTs placement in FPGA was not optimized by any special Quartus Prime tool, e.g. LogicLock, during compilation. Moreover, LUT propagation delay also depends on CMOS process parameters variation. However, the described phenomenon is beneficial in our case because broader set of different values inputs the combinatorial optimization and thus wider range of delays with fine granularity is covered.

A width of the output pulse varies from 4.9 ns to 42.5 ns and depends on input pulse width and provided delay. First PSC contains 15 LUTs and shrinks the pulse to the length of 4.3 ns corresponding to the delay introduced by these LUTs unless the width of the input pulse is shorter. Other PSCs contain only 4 LUTs and shrink the pulse to approximately 1 ns and are implemented before the last five (the largest) stages. When the signal passes only through the stages without PSCs the dead time increases linearly. If one of the last five stages with PSCs is used the dead time is a constant value which is given by the number of LUTs in the corresponding delay element.
FPGA output buffer provides pulses of +3.3V level which are converted by level shifter to +5V level. Level shifting contributes to zero-delay path with constant propagation delay which is between 1 ns to maximum of 6.7 ns in dependence on the load capacity \cite{SN74LVC4245A}. Output of the board is terminated by 50 Ohm resistor in series and lead out through SMA connector.


\section{Temporal performance of the delay line}
\label{Sec:temporal_characterization}

We characterized the following parameters of the developed DDL: linearity, granularity, jitter, and dead time. All measurements were performed using Lecroy 806Zi-B oscilloscope with a 6 GHz signal bandwidth and 40 GS/s sample rate in real-time mode. At least $5\times10^3$ samples were collected for each parameter measurement. Test pulses were generated by 240 MHz arbitrary waveform generator (Tektronix AFG3252) or 2 GHz clock generator (CG635 from Stanford Research) depending on the requirements of the particular measurement. To eliminate inter-channel jitter of the source, its output was split into two parts: one output was used as a signal pulse input for the delay generator, other one served as a reference for oscilloscope. Length of cables were selected, so that relative delay between reference pulse and signal pulse was zero. 

\begin{figure*}[ht!]
	\includegraphics[width=0.85\textwidth]{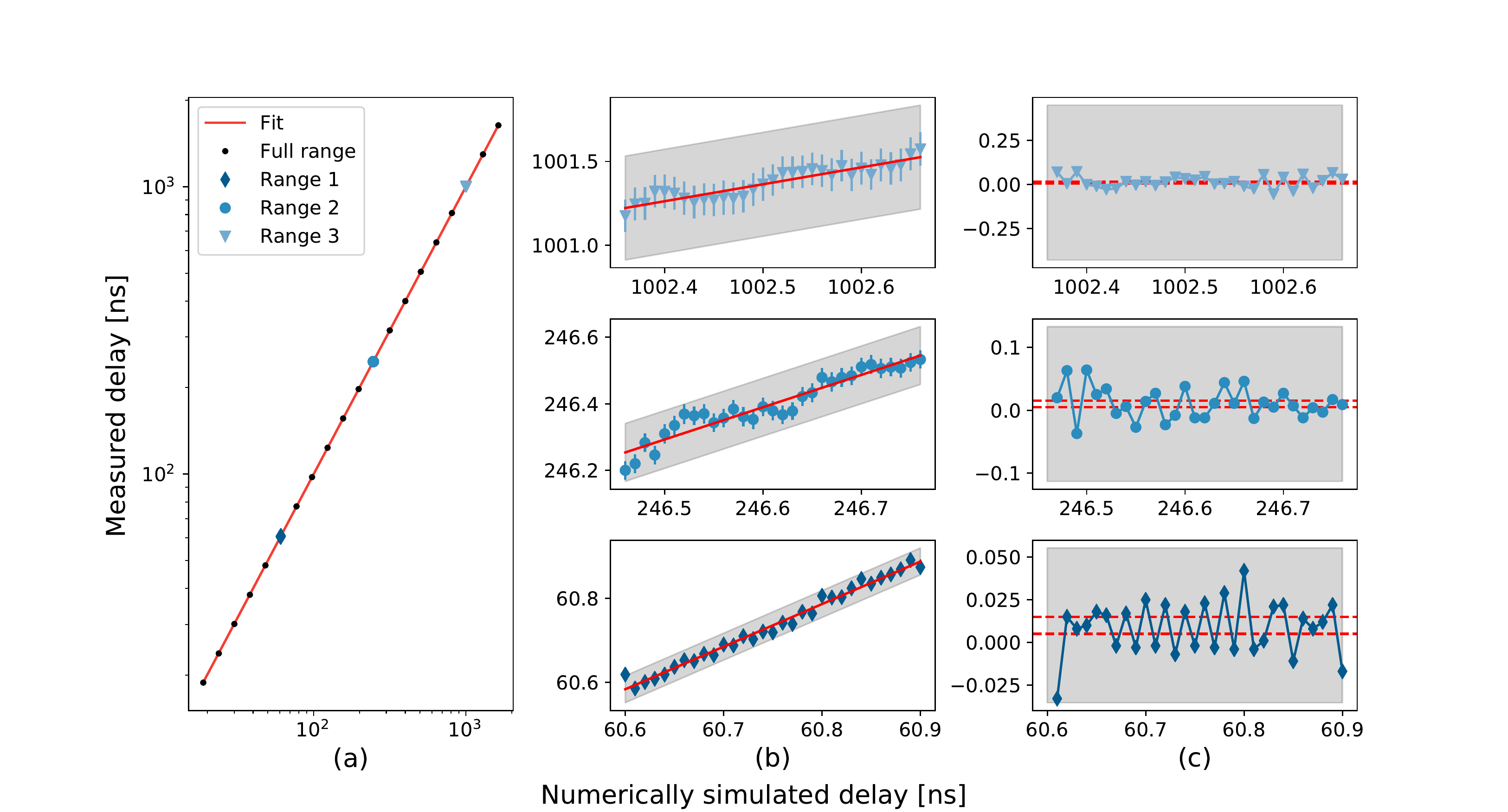}
	\caption{Linearity and granularity of the reported digital delay line. Panel (a) represents full dynamic range; panel (b) shows three particular time ranges to verify the linearity at the fine scale; red solid line stands for linear fit; panel (c) visualizes the granularity computed from the measured data shown in the panel (b). Red dashed lines represent the target granularity range of $10 \pm 5$ ps. Gray area shows $99.7\%$ confidence interval of the delay jitter.}
	\label{fig:linearity}
\end{figure*}

We directly measured the zero delay introduced by the device to be 18.808 ns for ambient temperature of $21.5\pm 0.2^{\circ}$C. Also, we performed the calibration of the DDL by measuring the delay of the individual stages resulting in the vector $(\tau_0, \ldots, \tau_{23})$.
In order to verify linearity of the DDL, we measured an actual delay introduced by the DDL when the corresponding target delay was set. We performed this measurement for 20 points uniformly distributed in log-scale, see Fig.~\ref{fig:linearity} (a). The results show linear dependence with unity slope. The upper limit of the generated delay has a value of 1635 ns.
Furthermore, we explored three particular ranges of the DDL output denoted by blue markers in Fig.~\ref{fig:linearity} (a) and compared the measured data with the linear fit, see Fig.~\ref{fig:linearity} (b). Small deviations from the perfect linear behavior visible in Fig.~\ref{fig:linearity} (b) are smaller than temporal jitter and will be discussed in the rest of this section.

Having measured the delay values in the three different regions of a full range of the DDL, we evaluate the differences of the measured consecutive delays to access the granularity, see Fig. \ref{fig:linearity} (c). The target granularity of $10(\pm 5)$ ps is shown as dashed red lines. The measured granularity fluctuates beyond the target region but it remains within $\pm100$ ps even for the largest delays. We offer several possible explanations of this effect. The main reason is the limited precision of the delay stages calibration caused by temporal jitter of the employed oscilloscope; the details are provided below. Furthermore, it is known that FPGAs are sensitive to process, voltage and temperature variations (PVT). Temperature undoubtedly affects DDL performance, see Section \ref{Sec:thermal_stability} for the details. The temperature was kept stable within 1 deg during linearity and granularity measurements.
FPGAs timing characteristics are influenced also by fluctuations of supply voltage. However, this effect can be neglected because voltage stabilizers were used to ensure sufficiently stable voltage with common laboratory power supply.

To measure the jitter, we utilize the same setup as for the linearity measurement. Instead of the mean value of the measured delay, the standard deviation of the delay (at 50\% signal level) was measured using statistical functions of the oscilloscope. The output jitter shows linear dependence on the delay; estimation from the data reads $\sigma_{\tau} [ps] = 4.726 + 0.098 \tau$, where $\sigma_{\tau}$ is the jitter and $\tau$ is the delay in nanoseconds, see Fig. \ref{fig:fig_3} \subref{fig:jitter}. The jitter growths linearly with the delay from 7 ps to 165 ps across the delay range, and causes uncertainty of the linearity measurements. Confidence intervals at the level of $99.7\%$ are shown as gray areas in Figs. \ref{fig:linearity} (b),(c). The observed granularity fluctuations are smaller than the temporal jitter.

The oscilloscope sample rate of 40 GS/s, equivalent to 25 ps per sample, limits the precision of the performed temporal characterization. The repeated measurements and their statistical processing allow to go beyond the sampling resolution. The inter-channel jitter of the oscilloscope at the level of 20 picoseconds seems to be the ultimate limitation of the precision of our characterization. Furthermore, the calibration precision is also affected, which can explain the observed fluctuations of the DDL granularity. Considering that the calibration errors of 24 stages are independent with the standard deviation $\delta = 20$ ps, the overall standard deviation of the maximum delay reaches  $\sqrt{24}\delta$. The peak-to-peak value uncertainty goes up to $6\sqrt{24}\delta = 588$ ps, which is larger than the observed fluctuation of the granularity.

By measuring the pulse width at the output of a FPGA delay line without PSCs we characterized the pulse spreading to be approximately
25 ps/LUT on average. However, due to inability to precisely predict the amount of pulse shrinking performed by PSCs, the overall dead time of the DDL caused by the pulse spreading cannot be computed easily. Instead, we performed a measurement of the maximum repetition frequency, which is transmitted through the DDL. The input square signal is provided by the clock generator. We used a counter (with 2.2~ns pulse-pair resolutions) to acquire the number of pulses at the output, and increased the repetition frequency of the input signal to the level when the output count rate is decreased to 95\% of the corresponding input repetition rate. Beyond this maximum frequency, the pulses of the input signal are spread to the whole period of the signal, and cannot be distinguished properly. The dead time of the DDL is defined as half of the period at the maximum frequency, see Fig. \ref{fig:fig_3} \subref{fig:deadtime}. The minimum measured dead time of the DDL is 4 ns; it is mainly due to the maximum input frequency of the used counter and propagation delay of the level shifter. The maximum dead time does not exceed 22.5 ns, i.e. 1.4\% of the maximum delay of the DDL.


\section{Thermal stability and calibration}
\label{Sec:thermal_stability}

To measure the performance of the developed DDL under varying thermal conditions, a box from thermo-isolating material (Styrofoam) was built. Temperature in the box was monitored by a certified thermometer with the precision of 0.15 deg. Each individual delay stage of the device was examined for thermal dependence in range from $30^{\circ}$C to $60^{\circ}$C with increment of $5^{\circ}$C. A weak linear dependence of the delay on temperature was revealed and will be discussed below.  

\begin{figure*}[ht!]
		\begin{subfigure}{0.85\columnwidth}
			\includegraphics[width=\columnwidth]{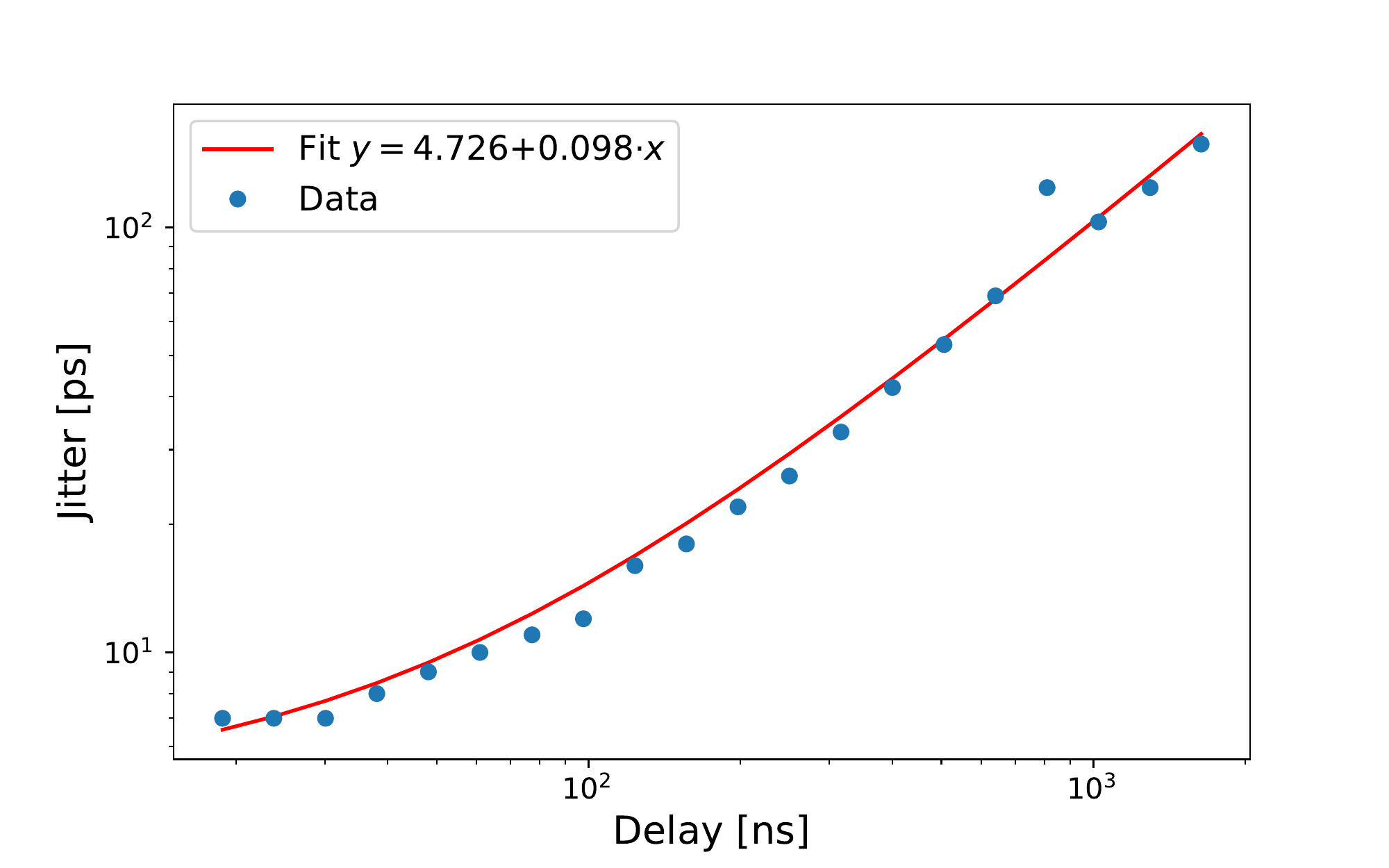}
			\caption{}
			\label{fig:jitter}
		\end{subfigure}
		\begin{subfigure}{0.85\columnwidth}
			\includegraphics[width=\columnwidth]{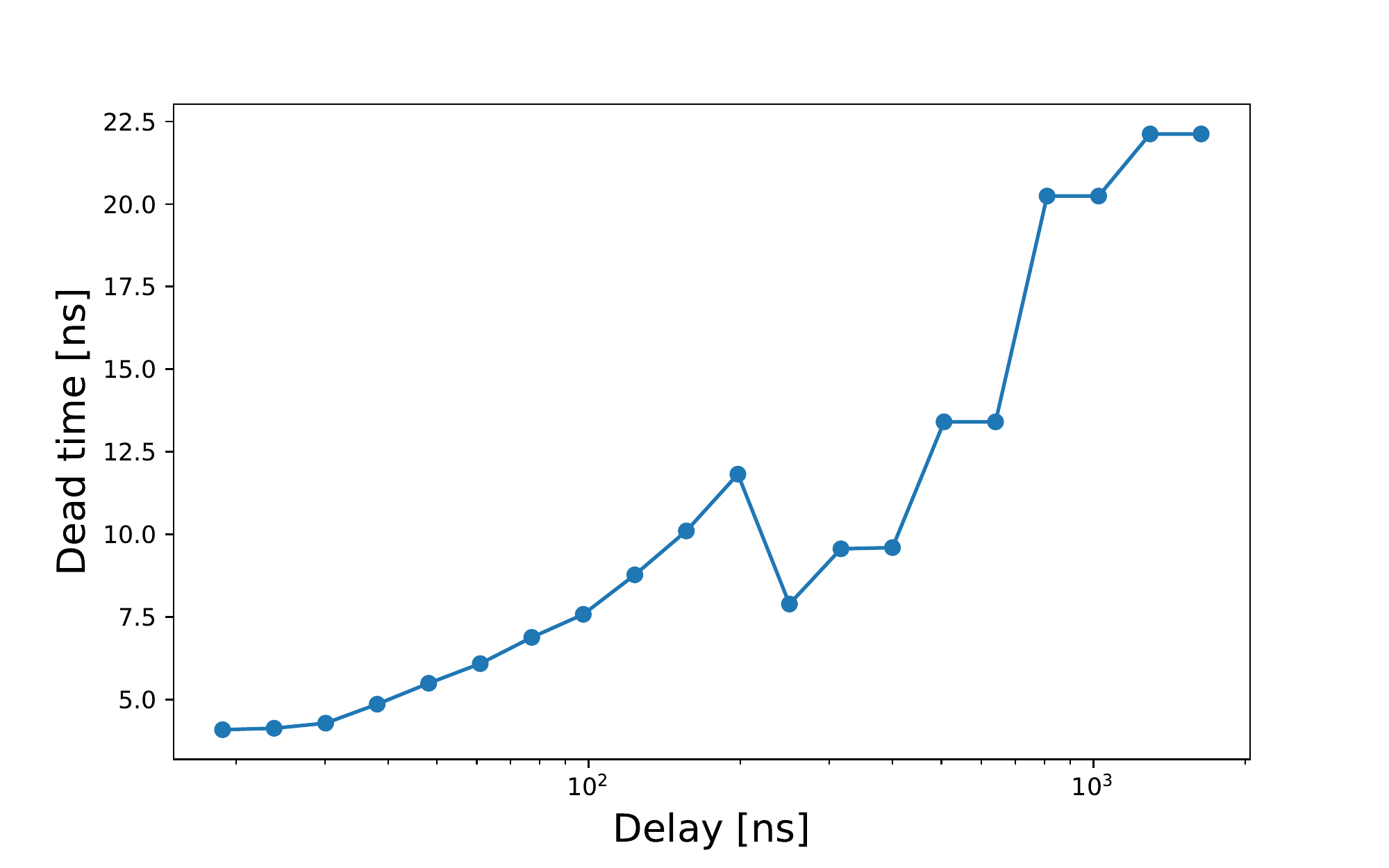}
			\caption{}
			\label{fig:deadtime}
		\end{subfigure}
		\begin{subfigure}{0.85\columnwidth}
			\includegraphics[width=\columnwidth]{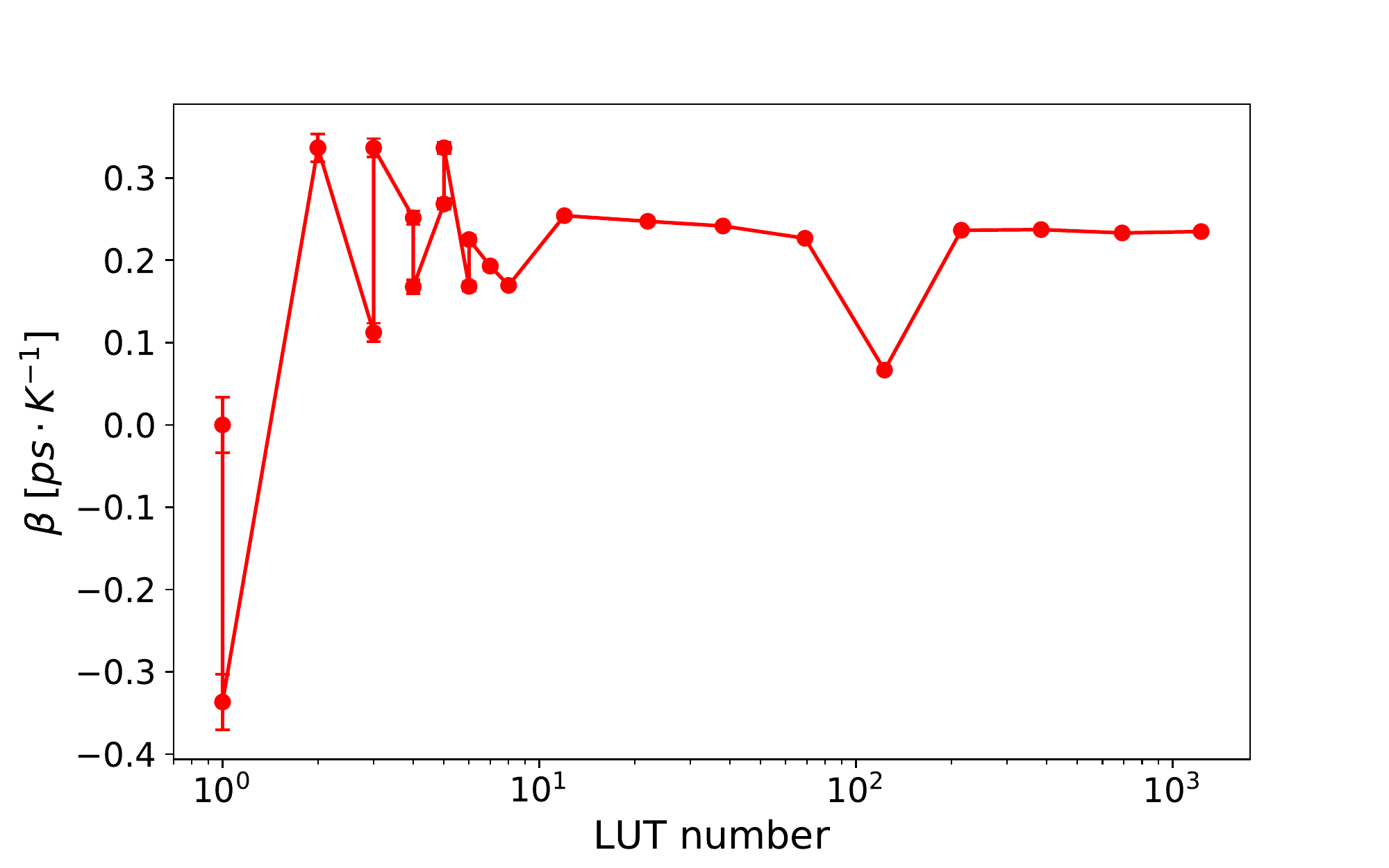}
			\caption{}
			\label{fig:beta_max_min_N}
		\end{subfigure}
			\begin{subfigure}{0.85\columnwidth}
			\includegraphics[width=\columnwidth]{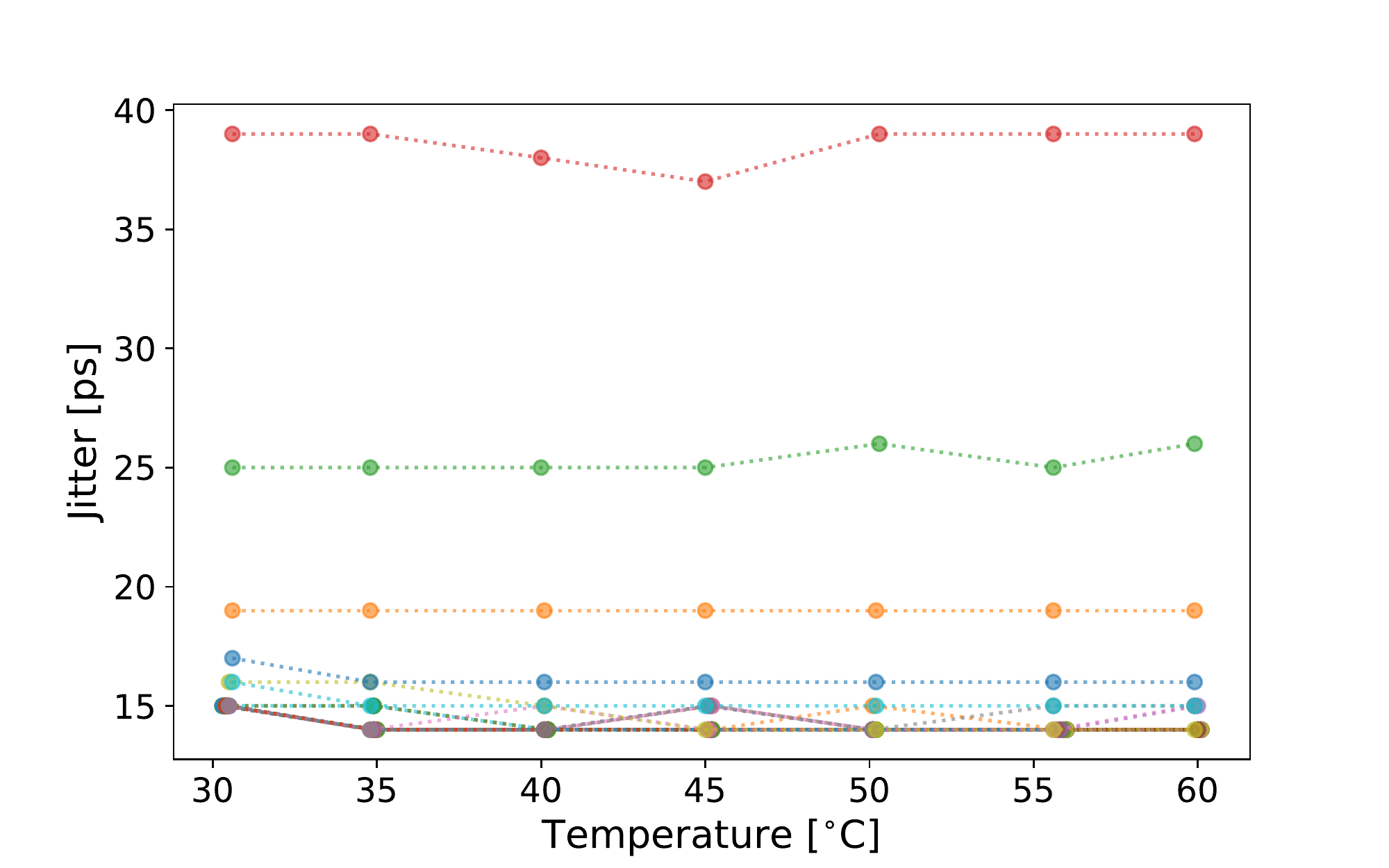}
			\caption{}
			\label{fig:jitter_temp}
		\end{subfigure}
	\caption{(a) Jitter dependence on the output delay. (b) Measured dead time of the delay line. Dead time growths linearly with the output delay up to value of $2\times 10^2$ ns. The stepwise behaviour of the remaining part of the curve is a direct consequence of pulse shrinking circuits. (c) Delay-temperature coefficient evolution over LUT number. (d) Jitter temperature dependence for all 24 stages (each line corresponds to individual stage)}
\label{fig:fig_3}
\end{figure*}

To verify the linear scaling of the delay time with the temperature, we calculated a delay-temperature coefficient defined as:
\begin{equation}
\beta = \frac{\tau_{max} - \tau_{min}}{(T_{max}-T_{min}) \cdot N}
\end{equation}
where $N$ is the number of LUTs used for the delay in each stage; $\tau_{max}$ and $\tau_{min}$ are the delay values corresponding to the maximum and minimum temperatures $T_{max}$ and $T_{min}$, respectively. Hence $\beta$ represents an average delay increase per Kelvin and a single LUT. In Fig. \ref{fig:fig_3} \subref{fig:beta_max_min_N} evolution of the $\beta$ over the LUT number can be seen. The values of $\beta$ obtained for small LUT numbers posses low confidence due to the measurement systematic errors. However, for larger number of LUTs, $\beta$ converges to its mean value of 0.2 ps $\cdot K^{-1}$ per LUT.
Having calculated $\beta$ one can upload its values into FPGA memory and perform real-time calibration of the delay generator configuration for the particular ambient temperature.

Furthermore, we characterize the temporal dependence of time jitter of the individual delay stages, see Fig. \ref{fig:fig_3}~\subref{fig:jitter_temp}. The measurement shows no significant changes in the jitter over $30^{\circ}$C.


\section{Possible improvements and extensions}

The issues of the reported delay generator are mainly the delay dependence on temperature, jitter and dead-time being functions of the delay. Here we propose how these characteristics can be improved.

Temperature stabilization of the whole device may significantly decrease delay dependence on temperature, although the proposed thermal calibration can solve this issue. Input voltage comparators, due to their thermal dependence are also contributing to overall DDL temperature dependence \cite{LTC6754}.

Utilizing latest generation FPGAs with finer CMOS/FET nm technologies may greatly contribute to the DDLs timing performance, mainly improving resolution and jitter characteristics \cite{Xia2019}. On the other hand, the reported device performance was primarily limited by our measurement setup precision.

To minimize the zero delay introduced by the DDL, special tools can be used during synthesis stage (Quartus LogicLock), allowing manual placement of the particular LUT blocks, placing them in the particular area of the FPGA chip as close to each other as possible, therefore minimizing propagation delay between individual blocks.


\section{Conclusion}

We report the novel design and implementation of the digital delay line based on non-blocking asynchronous circuit design. The digital circuitry was realized using FPGA logic primitives (LUTs) chained up into array blocks to form the set of corresponding delay stages. Calibration of these stages and selecting the optimum variations allow to cover a large range of output delays with fine granularity.
The presented device was built on custom board using Cyclone IV EP4CE6E22C6N FPGA. We also implemented pulse shrinking circuits to reduce negative effects of pulse dispersion. Furthermore, we performed a detailed experimental characterization of the delay line.
The obtained dynamic range of the generated delay reaches from 23 ns to 1635 ns. The target design granularity was $10\pm 5$ ps. The measured granularity deviates from the target range by tens of picoseconds, which is smaller than temporal jitter. We characterized the jitter dependence on the delay to be $\sigma_{\tau} [ps] = 4.726 + 0.098 \tau$, where $\sigma_{\tau}$ is the jitter and $\tau$ is the delay in nanoseconds. This corresponds to the jitter from 7 ps to 165 ps within the dynamical range of the device. We measured the device sensitivity to ambient temperature and estimated the delay-temperature coefficient $\beta = 0.2$ ps $\cdot$ K${}^{-1}$ per LUT. The developed delay line features dead time, which is a negligible fraction of the output delay. Particularly, the total dead time is 22.5 ns for the maximum delay of 1635 ns. This corresponds to frequency bandwidth of 20 MHz or larger within the dynamic range of the digital delay line, which is sufficient for delaying signals generated by majority of single-photon detectors. Despite the fact that FPGAs are typically not recommended for asynchronous methods in precise timing applications, we believe that this work will facilitate the development of digital delay lines utilizing non-blocking asynchronous design.



\acknowledgments
This work was supported by the Czech Science Foundation (project 19-19189S). MJ acknowledges \mbox{QuantERA} ERA-NET Cofund in Quantum Technologies, EU Horizon 2020, and Ministry of Education, Youth and Sports of Czech Republic (project HYPER-U-P-S, no. 8C18002).
GM also acknowledges the support by the Palack\'y University (project IGA-PrF-2020-004 and IGA-PrF-2021-002).

\section*{Data Availability}
The data that support the findings of this study are openly available in GitHub repository\cite{Mazin2021}.


\medskip

\bibliographystyle{aipnum4-1}
\bibliography{delay_biblio}

\begin{thebibliography}{28}%
\makeatletter
\providecommand \@ifxundefined [1]{%
 \@ifx{#1\undefined}
}%
\providecommand \@ifnum [1]{%
 \ifnum #1\expandafter \@firstoftwo
 \else \expandafter \@secondoftwo
 \fi
}%
\providecommand \@ifx [1]{%
 \ifx #1\expandafter \@firstoftwo
 \else \expandafter \@secondoftwo
 \fi
}%
\providecommand \natexlab [1]{#1}%
\providecommand \enquote  [1]{``#1''}%
\providecommand \bibnamefont  [1]{#1}%
\providecommand \bibfnamefont [1]{#1}%
\providecommand \citenamefont [1]{#1}%
\providecommand \href@noop [0]{\@secondoftwo}%
\providecommand \href [0]{\begingroup \@sanitize@url \@href}%
\providecommand \@href[1]{\@@startlink{#1}\@@href}%
\providecommand \@@href[1]{\endgroup#1\@@endlink}%
\providecommand \@sanitize@url [0]{\catcode `\\12\catcode `\$12\catcode
  `\&12\catcode `\#12\catcode `\^12\catcode `\_12\catcode `\%12\relax}%
\providecommand \@@startlink[1]{}%
\providecommand \@@endlink[0]{}%
\providecommand \url  [0]{\begingroup\@sanitize@url \@url }%
\providecommand \@url [1]{\endgroup\@href {#1}{\urlprefix }}%
\providecommand \urlprefix  [0]{URL }%
\providecommand \Eprint [0]{\href }%
\providecommand \doibase [0]{http://dx.doi.org/}%
\providecommand \selectlanguage [0]{\@gobble}%
\providecommand \bibinfo  [0]{\@secondoftwo}%
\providecommand \bibfield  [0]{\@secondoftwo}%
\providecommand \translation [1]{[#1]}%
\providecommand \BibitemOpen [0]{}%
\providecommand \bibitemStop [0]{}%
\providecommand \bibitemNoStop [0]{.\EOS\space}%
\providecommand \EOS [0]{\spacefactor3000\relax}%
\providecommand \BibitemShut  [1]{\csname bibitem#1\endcsname}%
\let\auto@bib@innerbib\@empty
\bibitem [{\citenamefont {Marques}\ and\ \citenamefont
  {Cruz}(2014)}]{Marques2014}%
  \BibitemOpen
  \bibfield  {author} {\bibinfo {author} {\bibfnamefont {J.}~\bibnamefont
  {Marques}}\ and\ \bibinfo {author} {\bibfnamefont {C.}~\bibnamefont {Cruz}},\
  }\href {\doibase https://doi.org/10.1016/j.nima.2014.01.060} {\bibfield
  {journal} {\bibinfo  {journal} {Nucl. Instrum. Methods Phys. Res. A}\
  }\textbf {\bibinfo {volume} {745}},\ \bibinfo {pages} {50 } (\bibinfo {year}
  {2014})}\BibitemShut {NoStop}%
\bibitem [{\citenamefont {Sliwczynski}\ \emph {et~al.}(2011)\citenamefont
  {Sliwczynski}, \citenamefont {Krehlik}, \citenamefont {Buczek},\ and\
  \citenamefont {Lipinski}}]{Sliwczynski2011}%
  \BibitemOpen
  \bibfield  {author} {\bibinfo {author} {\bibfnamefont {L.}~\bibnamefont
  {Sliwczynski}}, \bibinfo {author} {\bibfnamefont {P.}~\bibnamefont
  {Krehlik}}, \bibinfo {author} {\bibfnamefont {L.}~\bibnamefont {Buczek}}, \
  and\ \bibinfo {author} {\bibfnamefont {M.}~\bibnamefont {Lipinski}},\ }\href
  {\doibase 10.1109/tim.2010.2090696} {\bibfield  {journal} {\bibinfo
  {journal} {{IEEE} Trans. Instrum. Meas.}\ }\textbf {\bibinfo {volume} {60}},\
  \bibinfo {pages} {1480} (\bibinfo {year} {2011})}\BibitemShut {NoStop}%
\bibitem [{\citenamefont {Boaron}\ \emph {et~al.}(2018)\citenamefont {Boaron},
  \citenamefont {Korzh}, \citenamefont {Houlmann}, \citenamefont {Boso},
  \citenamefont {Rusca}, \citenamefont {Gray}, \citenamefont {Li},
  \citenamefont {Nolan}, \citenamefont {Martin},\ and\ \citenamefont
  {Zbinden}}]{Boaron2018}%
  \BibitemOpen
  \bibfield  {author} {\bibinfo {author} {\bibfnamefont {A.}~\bibnamefont
  {Boaron}}, \bibinfo {author} {\bibfnamefont {B.}~\bibnamefont {Korzh}},
  \bibinfo {author} {\bibfnamefont {R.}~\bibnamefont {Houlmann}}, \bibinfo
  {author} {\bibfnamefont {G.}~\bibnamefont {Boso}}, \bibinfo {author}
  {\bibfnamefont {D.}~\bibnamefont {Rusca}}, \bibinfo {author} {\bibfnamefont
  {S.}~\bibnamefont {Gray}}, \bibinfo {author} {\bibfnamefont {M.-J.}\
  \bibnamefont {Li}}, \bibinfo {author} {\bibfnamefont {D.}~\bibnamefont
  {Nolan}}, \bibinfo {author} {\bibfnamefont {A.}~\bibnamefont {Martin}}, \
  and\ \bibinfo {author} {\bibfnamefont {H.}~\bibnamefont {Zbinden}},\ }\href
  {\doibase 10.1063/1.5027030} {\bibfield  {journal} {\bibinfo  {journal}
  {Appl. Phys. Lett.}\ }\textbf {\bibinfo {volume} {112}},\ \bibinfo {pages}
  {171108} (\bibinfo {year} {2018})}\BibitemShut {NoStop}%
\bibitem [{\citenamefont {Flamini}, \citenamefont {Spagnolo},\ and\
  \citenamefont {Sciarrino}(2019)}]{Sciarrino2019}%
  \BibitemOpen
  \bibfield  {author} {\bibinfo {author} {\bibfnamefont {F.}~\bibnamefont
  {Flamini}}, \bibinfo {author} {\bibfnamefont {N.}~\bibnamefont {Spagnolo}}, \
  and\ \bibinfo {author} {\bibfnamefont {F.}~\bibnamefont {Sciarrino}},\
  }\href@noop {} {\bibfield  {journal} {\bibinfo  {journal} {Rep. Prog. Phys.}\
  }\textbf {\bibinfo {volume} {82}},\ \bibinfo {pages} {016001} (\bibinfo
  {year} {2019})}\BibitemShut {NoStop}%
\bibitem [{\citenamefont {Slussarenko}\ and\ \citenamefont
  {Pryde}(2019)}]{Pryde2019}%
  \BibitemOpen
  \bibfield  {author} {\bibinfo {author} {\bibfnamefont {S.}~\bibnamefont
  {Slussarenko}}\ and\ \bibinfo {author} {\bibfnamefont {G.~J.}\ \bibnamefont
  {Pryde}},\ }\href@noop {} {\bibfield  {journal} {\bibinfo  {journal} {Appl.
  Phys. Rev.}\ }\textbf {\bibinfo {volume} {6}},\ \bibinfo {pages} {041303}
  (\bibinfo {year} {2019})}\BibitemShut {NoStop}%
\bibitem [{\citenamefont {Zhang}\ \emph {et~al.}(2016)\citenamefont {Zhang},
  \citenamefont {Li}, \citenamefont {Hu}, \citenamefont {Yang}, \citenamefont
  {Jin},\ and\ \citenamefont {Jiang}}]{Zhang2016}%
  \BibitemOpen
  \bibfield  {author} {\bibinfo {author} {\bibfnamefont {C.}~\bibnamefont
  {Zhang}}, \bibinfo {author} {\bibfnamefont {W.}~\bibnamefont {Li}}, \bibinfo
  {author} {\bibfnamefont {Y.}~\bibnamefont {Hu}}, \bibinfo {author}
  {\bibfnamefont {T.}~\bibnamefont {Yang}}, \bibinfo {author} {\bibfnamefont
  {G.}~\bibnamefont {Jin}}, \ and\ \bibinfo {author} {\bibfnamefont
  {X.}~\bibnamefont {Jiang}},\ }\href {\doibase 10.1063/1.4967462} {\bibfield
  {journal} {\bibinfo  {journal} {Rev. Sci. Instrum.}\ }\textbf {\bibinfo
  {volume} {87}},\ \bibinfo {pages} {113107} (\bibinfo {year}
  {2016})}\BibitemShut {NoStop}%
\bibitem [{\citenamefont {Hlou{\v{s}}ek}\ \emph {et~al.}(2019)\citenamefont
  {Hlou{\v{s}}ek}, \citenamefont {Dudka}, \citenamefont {Straka},\ and\
  \citenamefont {Je{\v{z}}ek}}]{Hlousek2019}%
  \BibitemOpen
  \bibfield  {author} {\bibinfo {author} {\bibfnamefont {J.}~\bibnamefont
  {Hlou{\v{s}}ek}}, \bibinfo {author} {\bibfnamefont {M.}~\bibnamefont
  {Dudka}}, \bibinfo {author} {\bibfnamefont {I.}~\bibnamefont {Straka}}, \
  and\ \bibinfo {author} {\bibfnamefont {M.}~\bibnamefont {Je{\v{z}}ek}},\
  }\href {\doibase 10.1103/PhysRevLett.123.153604} {\bibfield  {journal}
  {\bibinfo  {journal} {Phys. Rev. Lett.}\ }\textbf {\bibinfo {volume} {123}},\
  \bibinfo {pages} {153604} (\bibinfo {year} {2019})}\BibitemShut {NoStop}%
\bibitem [{\citenamefont {Arabul}\ \emph {et~al.}(2020)\citenamefont {Arabul},
  \citenamefont {Paesani}, \citenamefont {Tancock}, \citenamefont {Rarity},\
  and\ \citenamefont {Dahnoun}}]{Arabul2020}%
  \BibitemOpen
  \bibfield  {author} {\bibinfo {author} {\bibfnamefont {E.}~\bibnamefont
  {Arabul}}, \bibinfo {author} {\bibfnamefont {S.}~\bibnamefont {Paesani}},
  \bibinfo {author} {\bibfnamefont {S.}~\bibnamefont {Tancock}}, \bibinfo
  {author} {\bibfnamefont {J.}~\bibnamefont {Rarity}}, \ and\ \bibinfo {author}
  {\bibfnamefont {N.}~\bibnamefont {Dahnoun}},\ }\href {\doibase
  10.1109/jphot.2020.2968724} {\bibfield  {journal} {\bibinfo  {journal}
  {{IEEE} Photon. J.}\ }\textbf {\bibinfo {volume} {12}},\ \bibinfo {pages} {1}
  (\bibinfo {year} {2020})}\BibitemShut {NoStop}%
\bibitem [{\citenamefont {Mendoza}\ \emph {et~al.}(2016)\citenamefont
  {Mendoza}, \citenamefont {Santagati}, \citenamefont {Munns}, \citenamefont
  {Hemsley}, \citenamefont {Piekarek}, \citenamefont {Mart{\'\i}n-L{\'o}pez},
  \citenamefont {Marshall}, \citenamefont {Bonneau}, \citenamefont {Thompson},\
  and\ \citenamefont {O’Brien}}]{OBrien2016}%
  \BibitemOpen
  \bibfield  {author} {\bibinfo {author} {\bibfnamefont {G.~J.}\ \bibnamefont
  {Mendoza}}, \bibinfo {author} {\bibfnamefont {R.}~\bibnamefont {Santagati}},
  \bibinfo {author} {\bibfnamefont {J.}~\bibnamefont {Munns}}, \bibinfo
  {author} {\bibfnamefont {E.}~\bibnamefont {Hemsley}}, \bibinfo {author}
  {\bibfnamefont {M.}~\bibnamefont {Piekarek}}, \bibinfo {author}
  {\bibfnamefont {E.}~\bibnamefont {Mart{\'\i}n-L{\'o}pez}}, \bibinfo {author}
  {\bibfnamefont {G.~D.}\ \bibnamefont {Marshall}}, \bibinfo {author}
  {\bibfnamefont {D.}~\bibnamefont {Bonneau}}, \bibinfo {author} {\bibfnamefont
  {M.~G.}\ \bibnamefont {Thompson}}, \ and\ \bibinfo {author} {\bibfnamefont
  {J.~L.}\ \bibnamefont {O’Brien}},\ }\href@noop {} {\bibfield  {journal}
  {\bibinfo  {journal} {Optica}\ }\textbf {\bibinfo {volume} {3}},\ \bibinfo
  {pages} {127} (\bibinfo {year} {2016})}\BibitemShut {NoStop}%
\bibitem [{\citenamefont {Xiong}\ \emph {et~al.}(2016)\citenamefont {Xiong},
  \citenamefont {Zhang}, \citenamefont {Liu}, \citenamefont {Collins},
  \citenamefont {Mahendra}, \citenamefont {Helt}, \citenamefont {Steel},
  \citenamefont {Choi}, \citenamefont {Chae}, \citenamefont {Leong},\ and\
  \citenamefont {Eggleton}}]{Xiong2016}%
  \BibitemOpen
  \bibfield  {author} {\bibinfo {author} {\bibfnamefont {C.}~\bibnamefont
  {Xiong}}, \bibinfo {author} {\bibfnamefont {X.}~\bibnamefont {Zhang}},
  \bibinfo {author} {\bibfnamefont {Z.}~\bibnamefont {Liu}}, \bibinfo {author}
  {\bibfnamefont {M.~J.}\ \bibnamefont {Collins}}, \bibinfo {author}
  {\bibfnamefont {A.}~\bibnamefont {Mahendra}}, \bibinfo {author}
  {\bibfnamefont {L.~G.}\ \bibnamefont {Helt}}, \bibinfo {author}
  {\bibfnamefont {M.~J.}\ \bibnamefont {Steel}}, \bibinfo {author}
  {\bibfnamefont {D.~Y.}\ \bibnamefont {Choi}}, \bibinfo {author}
  {\bibfnamefont {C.~J.}\ \bibnamefont {Chae}}, \bibinfo {author}
  {\bibfnamefont {P.~H.~W.}\ \bibnamefont {Leong}}, \ and\ \bibinfo {author}
  {\bibfnamefont {B.~J.}\ \bibnamefont {Eggleton}},\ }\href
  {https://doi.org/10.1038/ncomms10853} {\bibfield  {journal} {\bibinfo
  {journal} {Nat. Commun.}\ }\textbf {\bibinfo {volume} {7}} (\bibinfo {year}
  {2016})}\BibitemShut {NoStop}%
\bibitem [{\citenamefont {Kaneda}\ and\ \citenamefont
  {Kwiat}(2019)}]{Kaneda2019}%
  \BibitemOpen
  \bibfield  {author} {\bibinfo {author} {\bibfnamefont {F.}~\bibnamefont
  {Kaneda}}\ and\ \bibinfo {author} {\bibfnamefont {P.~G.}\ \bibnamefont
  {Kwiat}},\ }\href {\doibase 10.1126/sciadv.aaw8586} {\bibfield  {journal}
  {\bibinfo  {journal} {Sci. Adv.}\ }\textbf {\bibinfo {volume} {5}},\ \bibinfo
  {pages} {eaaw8586} (\bibinfo {year} {2019})}\BibitemShut {NoStop}%
\bibitem [{\citenamefont {\v{S}varc}\ \emph {et~al.}(2020)\citenamefont
  {\v{S}varc}, \citenamefont {Hlou\v{s}ek}, \citenamefont {Nov\'{a}kov\'{a}},
  \citenamefont {Fiur\'{a}\v{s}ek},\ and\ \citenamefont
  {Je\v{z}ek}}]{Svarc2020}%
  \BibitemOpen
  \bibfield  {author} {\bibinfo {author} {\bibfnamefont {V.}~\bibnamefont
  {\v{S}varc}}, \bibinfo {author} {\bibfnamefont {J.}~\bibnamefont
  {Hlou\v{s}ek}}, \bibinfo {author} {\bibfnamefont {M.}~\bibnamefont
  {Nov\'{a}kov\'{a}}}, \bibinfo {author} {\bibfnamefont {J.}~\bibnamefont
  {Fiur\'{a}\v{s}ek}}, \ and\ \bibinfo {author} {\bibfnamefont
  {M.}~\bibnamefont {Je\v{z}ek}},\ }\href {\doibase 10.1364/OE.385609}
  {\bibfield  {journal} {\bibinfo  {journal} {Opt. Express}\ }\textbf {\bibinfo
  {volume} {28}},\ \bibinfo {pages} {11634} (\bibinfo {year}
  {2020})}\BibitemShut {NoStop}%
\bibitem [{\citenamefont {Dudek}, \citenamefont {Szczepanski},\ and\
  \citenamefont {Hatfield}(2000)}]{Dudek2000}%
  \BibitemOpen
  \bibfield  {author} {\bibinfo {author} {\bibfnamefont {P.}~\bibnamefont
  {Dudek}}, \bibinfo {author} {\bibfnamefont {S.}~\bibnamefont {Szczepanski}},
  \ and\ \bibinfo {author} {\bibfnamefont {J.}~\bibnamefont {Hatfield}},\
  }\href {\doibase 10.1109/4.823449} {\bibfield  {journal} {\bibinfo  {journal}
  {IEEE J. Solid-State Circuits}\ }\textbf {\bibinfo {volume} {35}},\ \bibinfo
  {pages} {240–247} (\bibinfo {year} {2000})}\BibitemShut {NoStop}%
\bibitem [{\citenamefont {Cui}, \citenamefont {Li},\ and\ \citenamefont
  {Zhu}(2017)}]{Zhu2017}%
  \BibitemOpen
  \bibfield  {author} {\bibinfo {author} {\bibfnamefont {K.}~\bibnamefont
  {Cui}}, \bibinfo {author} {\bibfnamefont {X.}~\bibnamefont {Li}}, \ and\
  \bibinfo {author} {\bibfnamefont {R.}~\bibnamefont {Zhu}},\ }\href@noop {}
  {\bibfield  {journal} {\bibinfo  {journal} {Rev. Sci. Instrum.}\ }\textbf
  {\bibinfo {volume} {88}},\ \bibinfo {pages} {064703} (\bibinfo {year}
  {2017})}\BibitemShut {NoStop}%
\bibitem [{\citenamefont {B{\'{e}}dard}(1967)}]{Bedard1967}%
  \BibitemOpen
  \bibfield  {author} {\bibinfo {author} {\bibfnamefont {G.}~\bibnamefont
  {B{\'{e}}dard}},\ }\href@noop {} {\bibfield  {journal} {\bibinfo  {journal}
  {Proc. Phys. Soc.}\ }\textbf {\bibinfo {volume} {90}},\ \bibinfo {pages}
  {131} (\bibinfo {year} {1967})}\BibitemShut {NoStop}%
\bibitem [{\citenamefont {Straka}\ \emph {et~al.}(2020)\citenamefont {Straka},
  \citenamefont {Grygar}, \citenamefont {Hlou{\v{s}}ek},\ and\ \citenamefont
  {Je{\v{z}}ek}}]{Straka2020}%
  \BibitemOpen
  \bibfield  {author} {\bibinfo {author} {\bibfnamefont {I.}~\bibnamefont
  {Straka}}, \bibinfo {author} {\bibfnamefont {J.}~\bibnamefont {Grygar}},
  \bibinfo {author} {\bibfnamefont {J.}~\bibnamefont {Hlou{\v{s}}ek}}, \ and\
  \bibinfo {author} {\bibfnamefont {M.}~\bibnamefont {Je{\v{z}}ek}},\
  }\href@noop {} {\bibfield  {journal} {\bibinfo  {journal} {J. Light.
  Technol.}\ }\textbf {\bibinfo {volume} {38}},\ \bibinfo {pages} {4765}
  (\bibinfo {year} {2020})}\BibitemShut {NoStop}%
\bibitem [{\citenamefont {Jang}\ \emph {et~al.}(2007)\citenamefont {Jang},
  \citenamefont {Blieck}, \citenamefont {Veshapidze}, \citenamefont {Trachy},\
  and\ \citenamefont {DePaola}}]{Jang2007}%
  \BibitemOpen
  \bibfield  {author} {\bibinfo {author} {\bibfnamefont {H.~U.}\ \bibnamefont
  {Jang}}, \bibinfo {author} {\bibfnamefont {J.}~\bibnamefont {Blieck}},
  \bibinfo {author} {\bibfnamefont {G.}~\bibnamefont {Veshapidze}}, \bibinfo
  {author} {\bibfnamefont {M.~L.}\ \bibnamefont {Trachy}}, \ and\ \bibinfo
  {author} {\bibfnamefont {B.~D.}\ \bibnamefont {DePaola}},\ }\href {\doibase
  10.1063/1.2785029} {\bibfield  {journal} {\bibinfo  {journal} {Rev. Sci.
  Instrum.}\ }\textbf {\bibinfo {volume} {78}},\ \bibinfo {pages} {094702}
  (\bibinfo {year} {2007})}\BibitemShut {NoStop}%
\bibitem [{\citenamefont {Handa}, \citenamefont {Domalain},\ and\ \citenamefont
  {Kose}(2007)}]{Handa2007}%
  \BibitemOpen
  \bibfield  {author} {\bibinfo {author} {\bibfnamefont {S.}~\bibnamefont
  {Handa}}, \bibinfo {author} {\bibfnamefont {T.}~\bibnamefont {Domalain}}, \
  and\ \bibinfo {author} {\bibfnamefont {K.}~\bibnamefont {Kose}},\ }\href
  {\doibase 10.1063/1.2773636} {\bibfield  {journal} {\bibinfo  {journal} {Rev.
  Sci. Instrum.}\ }\textbf {\bibinfo {volume} {78}},\ \bibinfo {pages} {084705}
  (\bibinfo {year} {2007})}\BibitemShut {NoStop}%
\bibitem [{\citenamefont {Eyler}(2011)}]{Eyler2011}%
  \BibitemOpen
  \bibfield  {author} {\bibinfo {author} {\bibfnamefont {E.~E.}\ \bibnamefont
  {Eyler}},\ }\href {\doibase 10.1063/1.3523426} {\bibfield  {journal}
  {\bibinfo  {journal} {Rev. Sci. Instrum.}\ }\textbf {\bibinfo {volume}
  {82}},\ \bibinfo {pages} {013105} (\bibinfo {year} {2011})}\BibitemShut
  {NoStop}%
\bibitem [{\citenamefont {Hošák}\ and\ \citenamefont
  {Ježek}(2018)}]{Hosak2019}%
  \BibitemOpen
  \bibfield  {author} {\bibinfo {author} {\bibfnamefont {R.}~\bibnamefont
  {Hošák}}\ and\ \bibinfo {author} {\bibfnamefont {M.}~\bibnamefont
  {Ježek}},\ }\href {\doibase 10.1063/1.5019685} {\bibfield  {journal}
  {\bibinfo  {journal} {Rev. Sci. Instrum.}\ }\textbf {\bibinfo {volume}
  {89}},\ \bibinfo {pages} {045103} (\bibinfo {year} {2018})}\BibitemShut
  {NoStop}%
\bibitem [{\citenamefont {{Song}}\ \emph {et~al.}(2011)\citenamefont {{Song}},
  \citenamefont {{Liang}}, \citenamefont {{Zhou}}, \citenamefont {{Du}},
  \citenamefont {{Ma}},\ and\ \citenamefont {{Yue}}}]{Yue2011}%
  \BibitemOpen
  \bibfield  {author} {\bibinfo {author} {\bibfnamefont {Y.}~\bibnamefont
  {{Song}}}, \bibinfo {author} {\bibfnamefont {H.}~\bibnamefont {{Liang}}},
  \bibinfo {author} {\bibfnamefont {L.}~\bibnamefont {{Zhou}}}, \bibinfo
  {author} {\bibfnamefont {J.}~\bibnamefont {{Du}}}, \bibinfo {author}
  {\bibfnamefont {J.}~\bibnamefont {{Ma}}}, \ and\ \bibinfo {author}
  {\bibfnamefont {Z.}~\bibnamefont {{Yue}}},\ }in\ \href {\doibase
  10.1109/ICECC.2011.6067814} {\emph {\bibinfo {booktitle} {2011 International
  Conference on Electronics, Communications and Control (ICECC)}}}\ (\bibinfo
  {year} {2011})\ pp.\ \bibinfo {pages} {2116--2118}\BibitemShut {NoStop}%
\bibitem [{\citenamefont {Perko}\ and\ \citenamefont
  {Szplet}(2015)}]{Szplet2015}%
  \BibitemOpen
  \bibfield  {author} {\bibinfo {author} {\bibfnamefont {K.}~\bibnamefont
  {Perko}}\ and\ \bibinfo {author} {\bibfnamefont {R.}~\bibnamefont {Szplet}},\
  }\href@noop {} {\bibfield  {journal} {\bibinfo  {journal} {Meas. Autom.
  Monit.}\ }\textbf {\bibinfo {volume} {61}},\ \bibinfo {pages} {311} (\bibinfo
  {year} {2015})}\BibitemShut {NoStop}%
\bibitem [{\citenamefont {{Berrima}}, \citenamefont {{Blaquière}},\ and\
  \citenamefont {{Savaria}}(2017)}]{Savaria2017}%
  \BibitemOpen
  \bibfield  {author} {\bibinfo {author} {\bibfnamefont {S.}~\bibnamefont
  {{Berrima}}}, \bibinfo {author} {\bibfnamefont {Y.}~\bibnamefont
  {{Blaquière}}}, \ and\ \bibinfo {author} {\bibfnamefont {Y.}~\bibnamefont
  {{Savaria}}},\ }in\ \href {\doibase 10.1109/MWSCAS.2017.8053074} {\emph
  {\bibinfo {booktitle} {2017 IEEE 60th International Midwest Symposium on
  Circuits and Systems (MWSCAS)}}}\ (\bibinfo {year} {2017})\ pp.\ \bibinfo
  {pages} {918--921}\BibitemShut {NoStop}%
\bibitem [{\citenamefont {Zhang}\ \emph {et~al.}(2019)\citenamefont {Zhang},
  \citenamefont {Qin}, \citenamefont {Wang}, \citenamefont {Tong},
  \citenamefont {Rui}, \citenamefont {Rong},\ and\ \citenamefont
  {Du}}]{Zhang2019}%
  \BibitemOpen
  \bibfield  {author} {\bibinfo {author} {\bibfnamefont {W.-Z.}\ \bibnamefont
  {Zhang}}, \bibinfo {author} {\bibfnamefont {X.}~\bibnamefont {Qin}}, \bibinfo
  {author} {\bibfnamefont {L.}~\bibnamefont {Wang}}, \bibinfo {author}
  {\bibfnamefont {Y.}~\bibnamefont {Tong}}, \bibinfo {author} {\bibfnamefont
  {Y.}~\bibnamefont {Rui}}, \bibinfo {author} {\bibfnamefont {X.}~\bibnamefont
  {Rong}}, \ and\ \bibinfo {author} {\bibfnamefont {J.-F.}\ \bibnamefont
  {Du}},\ }\href {\doibase 10.1063/1.5119148} {\bibfield  {journal} {\bibinfo
  {journal} {Rev. Sci. Instrum.}\ }\textbf {\bibinfo {volume} {90}},\ \bibinfo
  {pages} {114702} (\bibinfo {year} {2019})}\BibitemShut {NoStop}%
\bibitem [{LTC(2015)}]{LTC6754}%
  \BibitemOpen
  \href
  {https://www.analog.com/media/en/technical-documentation/data-sheets/6754f.pdf}
  {\emph {\bibinfo {title} {High Speed Rail-to-Rail Input Comparator with LVDS
  Compatible Outputs}}},\ \bibinfo {organization} {Analog Devices} (\bibinfo
  {year} {2015})\BibitemShut {NoStop}%
\bibitem [{SN7(2015)}]{SN74LVC4245A}%
  \BibitemOpen
  \href {https://www.ti.com/lit/gpn/sn74lvc4245a} {\emph {\bibinfo {title}
  {Octal Bus Transceiver and {3.3V} to {5V} Shifter With 3-State Outputs}}},\
  \bibinfo {organization} {Texas Instruments} (\bibinfo {year}
  {2015})\BibitemShut {NoStop}%
\bibitem [{\citenamefont {Xia}, \citenamefont {Cao},\ and\ \citenamefont
  {Dong}(2019)}]{Xia2019}%
  \BibitemOpen
  \bibfield  {author} {\bibinfo {author} {\bibfnamefont {H.}~\bibnamefont
  {Xia}}, \bibinfo {author} {\bibfnamefont {G.}~\bibnamefont {Cao}}, \ and\
  \bibinfo {author} {\bibfnamefont {N.}~\bibnamefont {Dong}},\ }\href {\doibase
  10.1063/1.5084014} {\bibfield  {journal} {\bibinfo  {journal} {Rev. Sci.
  Instrum.}\ }\textbf {\bibinfo {volume} {90}},\ \bibinfo {pages} {044706}
  (\bibinfo {year} {2019})}\BibitemShut {NoStop}%
\bibitem [{\citenamefont {Mazin}()}]{Mazin2021}%
  \BibitemOpen
  \bibfield  {author} {\bibinfo {author} {\bibfnamefont {G.}~\bibnamefont
  {Mazin}},\ }\href@noop {} {}\bibinfo {howpublished}
  {\url{https://github.com/glebmmazin/delayline}},\ \bibinfo {note}
  {{N}on-blocking programmable delay line -- {GitHub} repository
  (2021)}\BibitemShut {NoStop}%
\end{thebibliography}%


\end{document}